\documentclass[fleqn,usenatbib]{mnras}

\usepackage{multirow}
\usepackage{threeparttable}

\usepackage{newtxtext,newtxmath}

\usepackage{txfonts}

\newcommand{\mJybm}{\mbox{mJy~beam${}^{-1}$}}

\DeclareRobustCommand{\VAN}[3]{#2}
\let\VANthebibliography\thebibliography
\def\thebibliography{\DeclareRobustCommand{\VAN}[3]{##3}\VANthebibliography}

\usepackage{graphicx}	
\usepackage{amsmath}	
\usepackage{amssymb}	

\title{Two Extreme Steep-Spectrum, Polarized Radio Sources Towards the Galactic Bulge}

\author[S. D. Hyman et al.]{
S. D. Hyman,$^{1}$\thanks{E-mail: shyman@sbc.edu}
D. A. Frail,$^{2}$
J. S. Deneva,$^{3}$
N. E. Kassim,$^{4}$
S. Giacintucci,$^{4}$
J. E. Kooi,$^{4}$
\newauthor
T. J. W. Lazio,$^{5}$
I. Joyner,$^{1}$
W. M. Peters,$^{4}$
V. Gajjar,$^{6}$
and A. P. V. Siemion$^{7}$
\\
$^{1}$Department of Engineering and Physics, Sweet Briar College, Sweet Briar, VA 24595, USA\\
$^{2}$National Radio Astronomy Observatory, P.O. Box O, Socorro, NM 87801, USA\\
$^{3}$George Mason University, resident at Space Science Division, Naval Research Laboratory, Washington, DC 20375, USA\\
$^{4}$Remote Sensing Division, Naval Research Laboratory, Washington, DC 20375, USA\\
$^{5}$Jet Propulsion Laboratory, California Institute of Technology, Pasadena, CA 91109\\
$^{6}$Department of Astronomy, University of California Berkeley, Berkeley CA 94720\\
$^{7}$SETI Institute, Mountain View, California 94043\\
}

\date{Accepted XXX. Received YYY; in original form ZZZ}

\pubyear{2021}

\begin{document}
\label{firstpage}
\pagerange{\pageref{firstpage}--\pageref{lastpage}}
\maketitle

\begin{abstract}

From an on-going survey of the Galactic bulge, we have discovered a number of compact, steep spectrum radio sources. In this present study we have carried out more detailed observations for two of these sources, located 43$^\prime$ and 12.7$^\circ$ from the Galactic Center. Both sources have a very steep spectrum ($\alpha\simeq{-3}$) and are compact, with upper limits on the angular size of 1-2$^{\prime\prime}$. Their flux densities appear to be relatively steady on timescales of years, months, and hours, with no indications of rapid variability or transient behavior. We detect significant circularly polarized emission from both sources, but only weak or upper limits on linear polarization. Neither source has a counterpart at other wavelengths and deep, high-frequency searches fail to find pulsations. We compare their source properties with other known compact, non-thermal source populations in the bulge (e.g. X-ray binaries, magnetars, the Burper, cataclysmic variables). Our existing data support the hypothesis that they are scatter broadened millisecond or recycled pulsars, either at the bulge or along the line of sight. We also consider the possibility that they may be a new population of Galactic radio sources which share similar properties as pulsars but lack pulsations; a hypothesis that can be tested by future large-scale synoptic surveys.


\end{abstract}


\section{Introduction}\label{sec:intro}

There has been a resurgence of interest in synoptic sky surveys at radio wavelengths \citep{2017A&A...598A.104S,2017MNRAS.464.1146H,2020PASP..132c5001L,2021MNRAS.502...60R}, driven technically by the development of sensitive wide-field instruments, and driven scientifically by the ability to explore new phase space, with hopes of discovering new and interesting astrophysical sources \citep[e.g.,][]{sfb+16, 2017MNRAS.466.1944M, bbm+18, 2019MNRAS.490.4898H}.
The Galactic center (GC) direction is the obvious place to begin any systematic exploration of the Galactic radio sky; lines of sight toward the GC can have as much as $100\times$ higher stellar densities than those toward high Galactic latitudes \citep[e.g.][]{1980ApJ...238L..17B,2002AcA....52..217U}. Imaging both the central parsec of the GC and the wider bulge region have yielded a treasure trove of new and interesting compact radio sources. Exploration of the time domain phase space has been particularly rewarding. Regular wide-field monitoring of the \hbox{GC}, for example, has led to the discovery of at least four radio-selected transients \citep{1992Sci...255.1538Z, 2002AJ....123.1497H, Hyman05, Hyman09}, and perhaps more \citep{2016ApJ...833...11C}. Deep multi-epoch imaging of the inner parsec of the GC has also revealed a large population of compact radio transients and variables \citep{2020ApJ...905..173Z}.


We have an on-going program to image the GC at radio wavelengths, building upon the early efforts of \citet{LaRosa00}, \citet{Nord04}, and \citet{2008ApJS..174..481L}. In our current GC work we have been focused on exploring the radio spectral domain, looking to identify compact, steep spectrum radio sources near the GC or within the bulge. Several different approaches have been taken to identify such sources, including imaging a 5 deg$^2$ region centered on Sgr\,A* \citep{2019ApJ...876...20H}, comparing existing large-scale meter and centimeter radio surveys \citep{2017MNRAS.468.2526B,2018MNRAS.474.5008D} and targeted imaging of the error ellipses of unidentified {\it Fermi} sources. Pulsars are the conventional  source population expected to be compact and steep spectrum \citep{blv13,mkk00}. Indeed, the motivation for searching {\it Fermi} error regions has been to identify the putative pulsars that are hypothesized to power the GC gamma-ray excess \citep[][and references therein]{2020PhRvD.102d3012A}. Nonetheless, in the course of carrying out these imaging searches we have identified an increasing number of compact, steep spectrum radio sources in the GC region that cannot readily be confirmed to be pulsars via detection of pulsations \citep{2019ApJ...876...20H,HymanInPrep}. While it is possible that these pulsations remain to be detected, evidence is growing that we may be seeing a new source population.

In this paper we present a detailed study of two of these compact, steep-spectrum sources from a sample of nearly two dozen such sources. In \S\ref{sec:Obs} we compile their properties (polarization, spectral index, variability, pulsations, and multi-wavelength counterparts) using archival data and new observations. In \S\ref{sec:conclusions} we summarize these results and then compare their source properties with other known compact, non-thermal sources populations in the bulge.

\section{Observations and Results}\label{sec:Obs}

The first of our two sources, C1748$-$2827, was reported by \citet{2019ApJ...876...20H}. With Galactic coordinates $(l,b)$=(0.69$^\circ$,$-0.22^\circ$), the source lies 43$^\prime$ from Sgr\,A*, or $\sim{11}^\prime$ from the massive molecular cloud Sgr\,B2, and it appears to lie in projection against a thin radio filament visible in a MeerKAT image of the GC \citep{2019Natur.573..235H}. In \citet{2019ApJ...876...20H}, we obtained a spectrum, an initial position, and a size constraint. We conducted initial pulsation searches, and placed a modest upper limit on the degree of polarization. In the observations described below we improve these parameters further, including deeper pulsation and single-pulse searches at higher frequencies, a detection of polarized emission, and we place limits on the short and long term variability.

The second source, C1709$-$3918, was found serendipitously outside the error circle of a {\it Fermi} $\gamma$-ray unidentified source \citep{HymanInPrep}. With Galactic coordinates $(l,b)$=(347.29$^\circ$,0.36$^\circ$), the source is 12.7$^\circ$ from the GC and lies 52$^\prime$ outside of the young shell-type supernova remnant RX\,J1713.7$-$3946 (G347.3$-$0.5). Below we present a complete set of new and archival observations on this compact, steep spectrum source. A log of the observations for both sources is given in Table~\ref{Obs Log}.

\begin{table*}
    \begin{minipage}{126mm}
	\centering
	\caption{Observations Log.}
	\smallskip
	\begin{threeparttable}
		\begin{tabular}{cccccc}
		\hline\hline \noalign{\smallskip} 
		\multirow{2}{*}{Date} & \multirow{2}{*}{Source} & Duration & Telescope & $\nu,\Delta\nu,\delta\nu$\tnote{a} & Temporal Res. \\
		 &    & (hrs)  & (Array)  & (MHz)  &  (sec) \\
		\hline\noalign{\smallskip}
        28 Jun 2008 & C1709$-$3918 & 0.67 & GMRT & 330, 21, 0.88 & 16.9  \\
        12 Mar 2009 & C1709$-$3918 & 4.2 & GMRT & 330, 21, 0.88 & 16.9  \\
        30 Jan 2015 & C1709$-$3918 & 0.5 & ATCA (6A) & 1400, 472, 1.0 & 10 \\
        27 Jun 2019 & C1709$-$3918 & 4.5 & uGMRT & 687, 264\tnote{b}, 0.98 & 5.4 \\
        15 Jan 2021 & C1748$-$2827 & 2.25 & VLA (A) & 1520, 840, 1.0 & 2 \\
        15 Jan 2021 & C1748$-$2827 & 2.25 & VLITE (A) & 341, 38, 0.1 & 2 \\
        25 Jan 2021 & C1748$-$2827 & 2.25 & VLA (A) & 1520, 840, 1.0 & 2 \\
        25 Jan 2021 & C1748$-$2827 & 2.25 & VLITE (A) & 341, 38, 0.1 & 2 \\
		\noalign{\smallskip} 
		\hline\noalign{\smallskip} 		
		\end{tabular}
	\medskip
		\begin{tablenotes}
		\footnotesize
		\item[a] Central frequency, effective bandwidth, and averaged channel width used in imaging.
		\item[b] A 264~MHz bandwidth could be used for Stokes I imaging, but for the polarization analysis only a 120~MHz-wide portion centered at 614~MHz contained useful data (see text).
		\end{tablenotes}	
	\end{threeparttable}
    \label{Obs Log}
    \end{minipage}
\end{table*}

\subsection{C1748--2827: VLA 1.5~GHz (L-Band)}\label{platinum_VLA_obs}

We observed the source C1748$-$2827 using the Karl G.~Jansky Very Large Array (VLA) of the National Radio Astronomy Observatory (NRAO) at L-band (1--2~GHz, VLA Project 20B-461).
We conducted full-polarization observations for a total time-on-source of 4.5~hours, separated over two days in 2021, January 15 and~25. For both days, the VLA was in the A configuration, its largest configuration, with an array diameter of $\approx35$ km. All 27 antennas were in operation. We used the default integration time (2~s).
We performed data calibration using the Common Astronomy Software Applications (CASA) data-reduction package \citep{2003ASSL..285..109G,2007ASPC..376..127M}. 

The 1 GHz bandwidth is divided into 16 bands (or spectral windows, spw), each with a resolution (or channel width) of 1~MHz. Due to radio frequency interference (RFI), predominately from GPS L1 downlink signals, spw~8 had to be excised entirely. We also removed the edge channels for each of these bands because the sensitivity drops off near the band edges; consequently, we retained 15 bands for analysis, each with an effective bandwidth of 56 MHz.

Observations of C1748$-$2827 were made in sets of three scans, each with a duration of five minutes. Each set of scans was bracketed by a scan of J1751$-$2524, the same complex gain and phase calibrator used in \cite{2019ApJ...876...20H}. We also observed the VLA primary flux density calibrator 3C\,286 to determine the parallel- and cross-hand delays, bandpass, absolute flux density scale, and absolute polarization angle. To calculate the instrumental polarization leakage (D-term) parameters, we observed J1407$+$2827, a low-polarization, steep spectrum source and one of the standard five primary low polarization leakage calibrators for the \hbox{VLA}. The amplitudes of the D-terms were $8-15$\%.

For each day of observation, we used the VLA calibration pipeline in CASA\footnote{%
\url{https://science.nrao.edu/facilities/vla/data-processing/pipeline}}
to perform the total intensity calibration and then calibrated the polarization manually, following the steps described in the VLA CASA guide.\footnote{\url{https://casaguides.nrao.edu/index.php?title=CASA\_Guides:Polarization\_Calibration\_based\_on\_CASA\_pipeline\_standard\_reduction:\_The\_radio\_galaxy\_3C75-CASA6.1.2.7-pipeline-2020.1.0.36}}
The VLA calibration pipeline performs basic flagging and calibration for Stokes~I continuum data.  The calibrated data for C1748$-$2827 for each day were then concatenated into one data measurement set for imaging and analysis.

Stokes~I images were made for each half of the 1 GHz-wide band using CASA's \texttt{tclean()} task.
We applied an inner $u$-$v$--plane cut of 10~k$\lambda$ in order to mitigate against confusion arising from large scale structure and to reduce the noise due to \hbox{RFI}. The Briggs weighting scheme was employed with the Robust parameter set to 0.5.
The resulting synthesized beam sizes (rms noise levels) are $2.79^{\prime\prime} \times 1.07^{\prime\prime}$ (17~$\mu$Jy~beam${}^{-1}$) and $2.05^{\prime\prime} \times 0.81^{\prime\prime}$ (13~$\mu$Jy~beam${}^{-1}$), respectively, for central frequencies of~1.27~GHz and~1.81~GHz. These images were exported and analyzed using the Astronomical Image Processing System (AIPS).
For each image, we fit C1748$-$2827 with a two-dimensional Gaussian component and a background level.
While possibly marginally resolved in the lower frequency image (1.27~GHz), C1748$-$2827 is unresolved in the higher frequency image (1.81~GHz).
We report here and in Table~\ref{Flux Densities} the peak flux densities of $0.62 \pm 0.02$ mJy~beam${}^{-1}$ and $0.20 \pm 0.01$ mJy~beam${}^{-1}$, respectively.
An image of the entire band yields a flux density of $0.37 \pm 0.01$ mJy~beam${}^{-1}$. The upper limit on the intrinsic diameter of the source is $0.9^{\prime\prime}$.

\begin{table*}
    \begin{minipage}{126mm}
	\centering
	\caption{Stokes I Detections.}
	\smallskip
	\begin{threeparttable}
		\begin{tabular}{cccccc}
		\hline\hline \noalign{\smallskip} 
		\multirow{2}{*}{Date} & \multirow{2}{*}{Source} & Frequency & RMS Noise\tnote{a} & Beam & $S_{peak}$\tnote{a,b}\\
		 &    & (GHz)  & ($\mJybm$)  & ($^{\prime\prime} \times ^{\prime\prime}$)  &  ($\mJybm$) \\
		\hline\noalign{\smallskip}
        28 Jun 2008 & C1709$-$3918 & 0.33 & 1.53 & $14.5 \times 7.0$ & $23.6 \pm 4.2$ \\
        12 Mar 2009 & C1709$-$3918 & 0.33 & 0.86 & $14.5 \times 8.6$ & $25.8 \pm 4.4$ \\
        27 Jun 2019 & C1709$-$3918 & 0.60 & 0.08  & $6.9 \times 4.0$  & $3.67 \pm 0.43$ \\
        27 Jun 2019 & C1709$-$3918 & 0.69 & 0.06 &  $6.0 \times 3.4$ & $2.38 \pm 0.28$ \\
        27 Jun 2019 & C1709$-$3918 & 0.77 & 0.05 &  $5.2 \times 3.2$ & $1.77 \pm 0.21$ \\
        15 + 25 Jan 2021 & C1748$-$2827 & 1.27 & 17 & $2.79 \times 1.07$ & $0.62 \pm 0.02$  \\
        15 + 25 Jan 2021 & C1748$-$2827 & 1.81 & 13 & $2.05 \times 0.81$ & $0.20 \pm 0.01$ \\
		\noalign{\smallskip} 
		\hline\noalign{\smallskip} 		
		\end{tabular}
	\medskip
		\begin{tablenotes}
		\footnotesize
		\item[a] C1709$-$3918 rms noise and flux densities include overall flux scale corrections (see text).
		\item[b] Peak flux densities are given since the sources are unresolved.
		\end{tablenotes}	
	\end{threeparttable}
    \label{Flux Densities}
    \end{minipage}
\end{table*}

The absolute flux scale was checked by comparing the flux density of a nearby field source, 2LC~000.735$-$0.261, to the 1.4~GHz value reported by \cite{2008ApJS..174..481L}.
The 2LC observations were made with the A-configuration \hbox{VLA}, and therefore match ours in resolution. While our flux density for this source is lower by 15\%, a reanalysis of the 2LC observation shows that the flux density uncertainty is of the same degree, and therefore could account for the difference. We find $\pm10\%$ consistency between the flux densities of C1748$-$2827 and six bright sources that are also located in the FOV of the 1.4~GHz VLA observation reported in \citet{2019ApJ...876...20H}. We adopt a 5\% uncertainty in the flux scale of the new observations. 

\begin{figure}
	\centering
	\includegraphics[width=\columnwidth, trim={40mm 40mm 40mm 40mm}, clip]{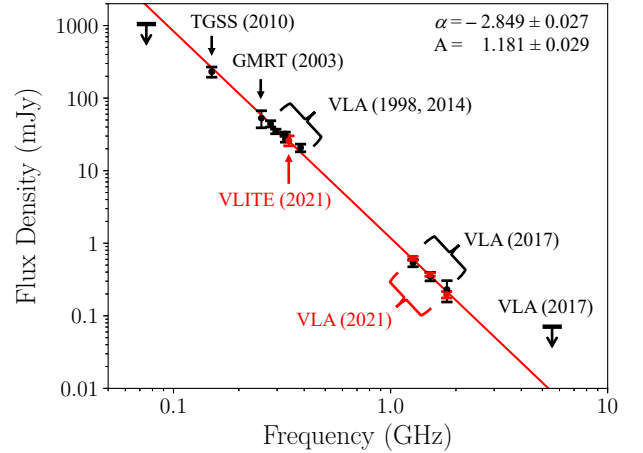}
	\caption{\small Radio spectrum of C1748$-$2827.  Results from this work [VLA (2021) and VLITE (2021)] are shown with red symbols, and results from previously reported data are shown with black symbols.  The red line is a power-law fit ($S$=$A$[$\nu$/1 GHz]$^{-\alpha}$); values for $A$ and $\alpha$ are reported in the upper right corner. The GMRT (2003), TGSS (2010), VLA (1998), VLA (2014 and 2017) data, and the upper limit at 74 MHz are from \citet{roy13}, \citet{ijmf16}, \citet{Nord04}, \citet{2019ApJ...876...20H}, and \citet{Brogan03}, respectively.}
	 \label{fig:platinum_spectra}
\end{figure}

Data from a 40~MHz band centered at 340 MHz were recorded simultaneously with our new 15 and 25 January 2021 1.5~GHz observations by the VLA Low-band Ionosphere and Transient Experiment (VLITE) \citep{2016SPIE.9906E..5BC}. Primary calibration and data editing were performed by the dedicated VLITE automated pipeline \citep{2016ApJ...832...60P}. The observations were combined and imaged with baselines shorter than 2.0 k$\lambda$ removed. The final image has an rms noise of 1.5 mJy~beam${}^{-1}$. We report a peak flux density of $26.1 \pm 4.1$ mJy~beam${}^{-1}$, which includes a $15\%$ absolute calibration uncertainty. Figure~\ref{fig:platinum_spectra} shows the spectrum of C1748$-$2827 with the new flux densities, updated from that reported in \citet{2019ApJ...876...20H}. 

A weighted fit results in a spectral index of $\alpha$ = $-2.85 \pm 0.03$. The fitted position of C1748$-$2827 on the higher resolution 1.81 GHz image is: (J2000) RA 17:48:07.070 $\pm 0.015 s$, DEC $-$28:27:41.68 $\pm 0.1^{\prime\prime}$, where the uncertainties are based on a comparison of the positions of several field sources to their 2LC positions. The position, spectral index and angular size constraints are consistent with \citet{2019ApJ...876...20H}, but with slightly improved uncertainties.


\subsubsection{C1748$-$2827: Polarization Imaging and Results}\label{platinum_VLA_pol}

We generated Stokes~\hbox{Q}, \hbox{U}, and~V images for the lower half of the 1~GHz-wide L~band using the CASA task \texttt{tclean()}. As for the Stokes~I imaging described above, we enforced an inner $u$-$v$--plane cut of~10~k$\lambda$. 
Natural weighting was used in the imaging in order to obtain a higher sensitivity, as this would allow detection of fainter polarized emission, instead of using uniform weighting that maximizes the angular resolution at the cost of sensitivity.
We also generated a set of corresponding Stokes~I maps using natural weighting in order to directly compare Stokes~\hbox{Q}, \hbox{U}, and~V flux densities; consequently, all fractional polarizations we present for this source are derived from maps with natural weighting.

\begin{table*}
    \begin{minipage}{126mm}
	\centering
	\caption{Stokes~\hbox{I}, \hbox{Q}, \hbox{U}, and~V peak flux densities and the fractional circular polarization for the radio sources C1748$-$2827 and C1709$-$3918.}
	\smallskip
	\begin{threeparttable}
		\begin{tabular}{ccccccc}
		\hline\hline \noalign{\smallskip} 
		\multirow{2}{*}{Date} & \multirow{2}{*}{Source} & I\tnote{a} & Q & U & V & \multirow{2}{*}{V/I} \\
		 &    & ($\mu$Jy beam$^{-1}$)  & ($\mu$Jy beam$^{-1}$)  & ($\mu$Jy beam$^{-1}$)  &  ($\mu$Jy beam$^{-1}$) &\\
		\hline\noalign{\smallskip}
        15 Jan 2021\tnote{b} &C1748$-$2827 &$610\pm20$ & $|80|$\tnote{c} & $|80|$\tnote{c} & $-88\pm16$  & $14.4\pm2.7\%$\\
        25 Jan 2021\tnote{b} &C1748$-$2827 &$590\pm13$ & $|50|$\tnote{c} & $|50|$\tnote{c} & $-83\pm10$  & $14.1\pm1.7\%$\\
        15 + 25 Jan 2021\tnote{b} & C1748$-$2827    &$596\pm12$ & $|45|$\tnote{c} & $|45|$\tnote{c}     & $-86\pm9$ & $14.4\pm1.5\%$ \\
        15 + 25 Jan 2021\tnote{d}  & C1748$-$2827   &$190\pm11$ & $|40|$\tnote{c} & $|40|$\tnote{c}     & $-27\pm8$ & $14.2\pm4.3\%$ \\
    \hline
        27 Jun 2019\tnote{e} & C1709$-$3918 &  $3200\pm90$ & $+237\pm47$ & $|235|$\tnote{c} & $-440\pm20$ & $13.8\pm0.9\%$ \\
		\noalign{\smallskip} 
		\hline\noalign{\smallskip} 		
		\end{tabular}
	\medskip
		\begin{tablenotes}
		\footnotesize
		\item[a] Determined from Stokes I maps generated using the same parameters as the Q, U, and V maps for direct comparison.
        \item[b] Center frequency (bandwidth) 1.26 GHz (448 MHz).
        \item[c] $5\sigma$ upper limit.
        \item[d] Center frequency (bandwidth) 1.84 GHz (384 MHz).
        \item[e] Center frequency (bandwidth) for Stokes I and V is 618 MHz (120~MHz); for Stokes Q and U is 633.5 MHz (40~MHz). Peak flux densities do not reflect a $+10\%$ flux scale correction.
		\end{tablenotes}	
	\end{threeparttable}
    \label{tab:IQUVflux}
    \end{minipage}
\end{table*}

The resulting synthesized beam size for the I, Q, U, and V maps was $3.29^{\prime\prime} \times 1.44^{\prime\prime}$ for a central frequency of 1.26 GHz. The rms noise levels for the I, Q, U, and V maps were 12, 9, 9, and $9\,\mu$Jy beam$^{-1}$, respectively. The peak flux densities for each Stokes parameter is given in Table~\ref{tab:IQUVflux}, for the combined data set as well as for each day of observation (15 and 25 January 2021). No linear polarization was detected in maps generated using the entire $\approx500$ MHz bandwidth provided by spw $0-7$ nor in the individual spw (each with $\approx56$ MHz bandwidth). The $5\sigma$ upper limit placed on the total linear polarization ($P\equiv\sqrt{Q^2+U^2}$) is $45\,\mu$Jy beam$^{-1}$($7.6\%$ fractional polarization) using the combined data. The rms errors reported here are determined from a propagation of radiometer noise errors. These new data are a substantial improvement over the early linear and circular polarization limits of 0.4 $\mJybm$ and 0.3 $\mJybm$, respectively \citep{2019ApJ...876...20H}.

\begin{figure}
	\centering
	\includegraphics[width=\columnwidth, trim = {40mm 50mm 50mm 50mm}, clip]{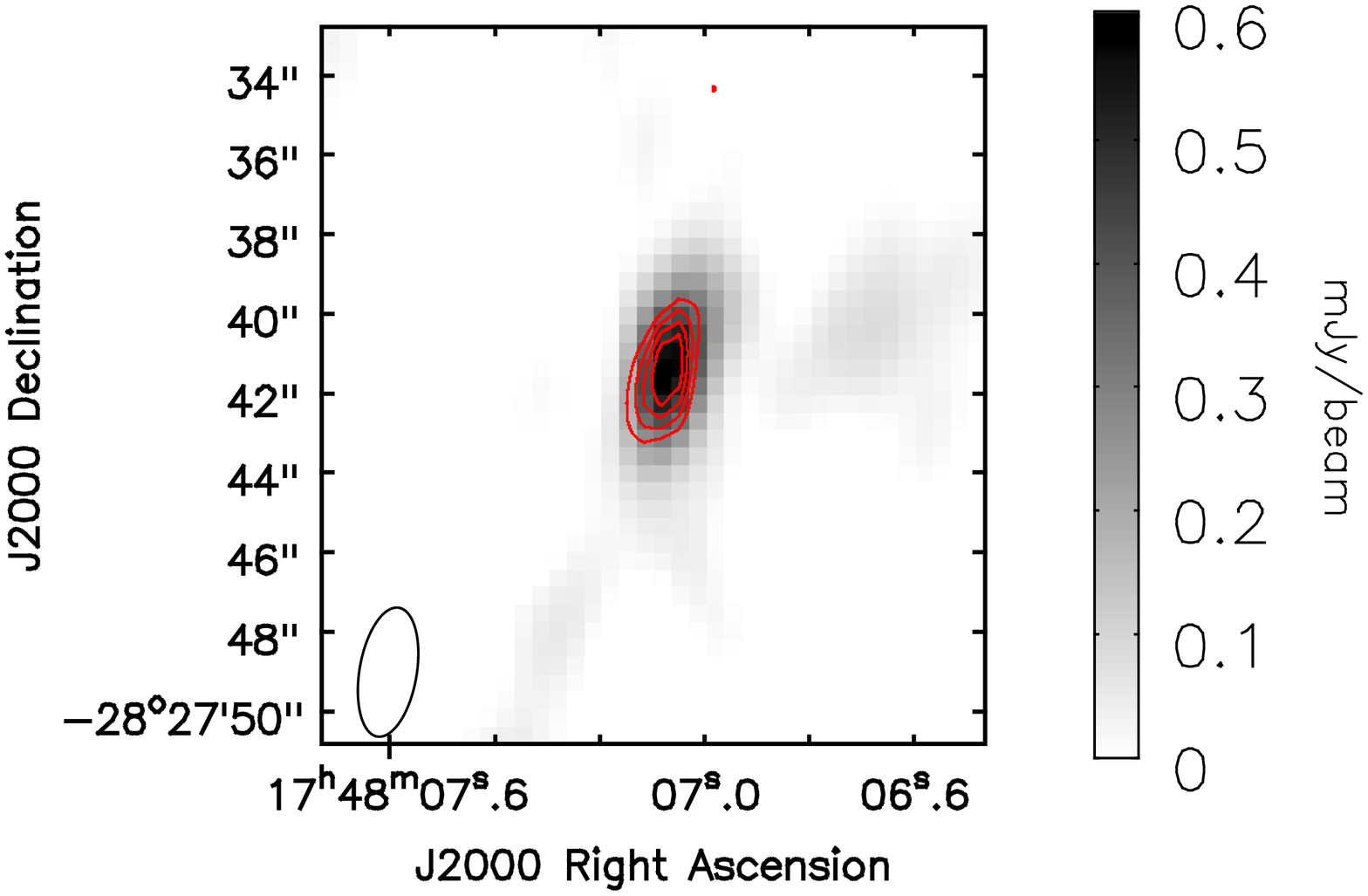}
	\caption{\small Clean map of the total intensity and circular polarization structure of C1748$-$2827 using the combined observations from 2021 January 15 and 25. This image is generated using the lower half of the $1-2$ GHz bandpass centered at a frequency of 1.26 GHz. Grayscale indicates the magnitude of the total intensity (Stokes I). Contours show the circular polarization intensity (Stokes V) plotted at $+5$, $-5$, $-6$, $-7$, and $-8$ times the rms error, $9\,\mu$Jy beam$^{-1}$ (positive and negative contours are plotted with red dashed and solid lines, respectively).  The resolution of the image is $3.29^{\prime\prime} \times 1.44^{\prime\prime}$ (shown in the lower-left corner).  The fractional circular polarization is $14.4\pm1.5\%$.}
	 \label{fig:radio_map_platinum}
\end{figure}

Circular polarization (Stokes V), though, was clearly detected. Figure~\ref{fig:radio_map_platinum} shows both the circular polarization intensity and total intensity structure for this source. The fractional circular polarization for the combined data is $14.4\pm1.5\%$. Dividing the data into 56 MHz-wide subbands demonstrates that the fractional polarization is relatively consistent across this band: $18.3\pm3.1\%$ (1.039 GHz), $11.9\pm3.2\%$ (1.103 GHz), $16.0\pm4.9\%$ (1.167 GHz), $8.9\pm4.5\%$ (1.231 GHz), $25.2\pm6.7\%$ (1.295 GHz), $21.8\pm5.1\%$ (1.359 GHz), $19.1\pm3.9\%$ (1.423 GHz), and $11.0\pm4.1\%$ (1.487 GHz). Table~\ref{tab:IQUVflux} also demonstrates that the Stokes I and V fluxes of C1748$-$2827 are consistent between the two days of observation.

We focused our analysis on the lower half of the L-band observations because the fractional circular polarization is $14.4\%$; therefore, due to the steep spectrum, Stokes V is much more difficult to detect at higher frequencies. However, we did generate full Stokes images for the upper half of the 1~GHz-wide L~band using the same imaging parameters described above, with results reported in Table~\ref{tab:IQUVflux}. A faint Stokes~\hbox{V} signal was marginally detected at the $3.4\sigma$-level with a flux density of $-27\pm8\,\mu\mathrm{Jy}$ at a center frequency of 1.839 GHz, implying a fractional circular polarization of $14.2\pm4.3\%$ at these higher frequencies. While the usable bandwidth is smaller (384 MHz in the upper band compared to 448 MHz in the lower band), there is slightly less radiometer noise in the upper band because there is much less RFI. The results from the lower and upper band Stokes~\hbox{V} data suggest that the fractional circular polarization is consistent across the full 1-GHz bandwidth.

\subsection{C1709--3918: GMRT 330 MHz (P-Band)}\label{Isab Pband}

As noted earlier, C1709$-$3918 was initially found serendipitously outside the error circle of a {\it Fermi} $\gamma$-ray unidentified source. The source appeared noteworthy since it was bright in archival 150 MHz observations made by the Giant Metrewave Radio Telescope (GMRT) for the TIFR GMRT Sky Survey (TGSS) \citep{ijmf16}, but not visible in a 1.5 GHz VLA image \citep{HymanInPrep}. This suggested either that C1709$-$3918 had a steep spectrum or it was highly variable. To better clarify its properties, we searched archival data and made new observations.

The source is located within the FOV of two GMRT archival 330 MHz observations (projects 14SRA01 and 15FAA01) at a distance of,  respectively, $3^{\prime}$ and $51^{\prime}$ from the phase center.
The 14SRA01 observation targeted the source 347.25+.375 (RA, DEC = 17:09:45, $-$39:19:42) and was carried out in June 2008 for a total of $\sim 40$ minutes on source. The 15FAA01 observation (March 2009) was pointed on G347.3-0.5 (RA, DEC = 17:13:36, $-$39:46:28) for a total of $4.2$ hours on target. Both projects used pre-upgrade hardware 2-sideband correlator that generated 128 frequency channels across each of the two 16 MHz-wide upper-side and lower-side bands (USB and LSB).
Observational parameters are listed in Table~\ref{Obs Log}.

For each observation, the USB and LSB data were reduced separately in AIPS. Initial bandpass and gain calibrations were derived using the standard primary calibrators 3C\,48 and 3C\,286, whose flux density models were set using the \citet{2012MNRAS.423L..30S} scale. The sources 1714--252 (14SRA01), 1712--281 and 1830--360 (15FAA01) were included in the observations and used to calibrate the data in phase. After initial calibration, the task RFLAG was used to excise visibilities affected by RFI, followed by manual flagging to remove residual bad data. The calibration was then repeated on each edited dataset, and the calibrated target visibilities were extracted from the multi-source files, averaged to 12 channels, each 0.875 MHz wide. A number of phase self-calibration iterations were applied to the USB and LSB data individually, using wide-field imaging and decomposing the primary beam area into a large number of smaller facets. 
For the 14SRA01 observation, the final USB and LSB images were obtained using only baselines longer than 2 k$\lambda$ and restored with a common beam of $14^{\prime\prime}.50 \times 6^{\prime\prime}.96$. The images were then combined together, achieving an average rms noise of $\sim0.9$ $\mJybm$. Following a similar procedure, we obtained a final combined image for the 15FAA01 observation with a beam of $14^{\prime\prime}.46 \times 8^{\prime\prime}.56$ and average rms of $\sim0.1$ $\mJybm$. Finally, we used PBCOR\footnote{\url{http://www.ncra.tifr.res.in:8081/\textasciitilde ngk/primarybeam/beam.html}} to
correct both images for the GMRT primary beam response.

C1709$-$3918 is detected and unresolved on both images. To check the accuracy of the flux scale, we compared the primary-beam corrected fluxes for a number of compact sources in the field to their predicted values based on their TGSS (150 MHz) and NVSS (1.4 GHz) flux densities \citep{1998AJ....115.1693C}. For 14SRA01, we used 6 sources and derived a flux scale correction of 1.7. Applying this correction, we obtained a flux density of
$23.6\pm4.2$ $\mJybm$ for C1709$-$3918, also listed in Table~\ref{Flux Densities}. The errors include local image rms, 8$\%$ absolute flux calibration uncertainty (e.g., Chandra, Ray \& Bhatnagar 2004), and 15\% flux scale correction uncertainty. 
For the longer observation 15FAA01, we estimated a flux scale correction of 6 based on TGSS/NVSS comparison for 3 field sources, and obtained a similar flux density of $25.8\pm4.4$ $\mJybm$, with errors accounting for local rms, calibration (8\%) and flux scale correction uncertainty (15\%).





\subsection{C1709-3918: uGMRT Band 4}\label{isabel1_GMRT_obs}

We observed C1709$-$3918 with the uGMRT (upgraded GMRT) in Band 4 (550-950 MHz) on 27 June 2019 for a total of 7 hours, including calibration overheads (project 36044).
The observations were carried out in full polarimetric mode with 27 working antennas. The data were collected in spectral-line mode using the wide-band backend with a total bandwidth of 400 MHz, 8192 frequency channels, and 5.4 second integration time.
We observed 3C\,48 and 3C\,286 at the beginning and end of the observation as delay, bandpass and absolute flux density calibrators. 3C\,286 was 
also selected as standard calibrator for the absolute polarization angle. The source 1714$-$252 was used as phase calibrator, and 1407+384 was 
observed as leakage term calibrator to determine the instrumental polarization.

The data were reduced using a combination of CASA and AIPS.
For the non-polarization calibration in CASA, we followed a standard procedure as detailed, for instance, in the GMRT Radio Astronomy School 2019 CASA tutorial\footnote{{http://www.ncra.tifr.res.in/~ruta/ras2019/CASA-tutorial.html}}. The flux density scale was set using Perley \& Butler (2017). 
Data affected by RFI were excised using AOFLAGGER (Offringa et al. 2012). After
calibration and flagging, the usable bandwidth decreased to 264 MHz with a central frequency of 687 MHz. The target visibilities 
were then divided into five roughly equal parts, imported into AIPS, and self-calibrated in phase individually. Finally, the self-calibrated 
data sets were combined back together and final images were produced with the task IMAGR with an inner uv-plane cut of $>$ 2 k$\lambda$. Images were made for each of three 88 MHz-wide portions of the band with central frequencies of 599, 687, and 774 MHz. The rms noise and resolution of the 687 MHz image are $\sim 70$ $\mu$Jy beam$^{-1}$ and $6^{\prime\prime}.0 \times 3^{\prime\prime}.4$. 

The flux densities of C1709$-$3918 and field sources were measured by fitting a Gaussian plus background, and correcting for the GMRT primary beam response\footnote{\url{http://www.ncra.tifr.res.in/ncra/gmrt/gmrt-users/observing-help/ugmrt-primary-beam-shape}}, using the task JMFIT in AIPS. 
The source is unresolved on each of the images. The fitted position of the source is 17:09:55.570 $\pm$ 0.035s, $-$39:18:02.05 $\pm$ $0.61^{\prime\prime}$, consistent across the three images. The JMFIT upper limit on the deconvolved size is $2.0^{\prime\prime}$. The position of a bright field source agrees with the known position of NVSS J170835$-$392935 to within the RA ($\pm 0.05s$) and DEC ($\pm 0.7^{\prime\prime}$) uncertainties of the latter. Following the results of \citet{2008MNRAS.383...75G}, who compared the scatter in position estimates of sources found in their narrow band GMRT 610 MHz survey to those from the 1.4 GHz VLA FIRST survey \citep{1995ApJ...450..559B}, the position error above includes their systematic errors of $0.4^{\prime\prime}$ (RA) and $0.6^{\prime\prime}$ (DEC) added in quadrature with the fitting error. We caution that there may be un-modeled offsets in the uGMRT 610 MHz (Band 4) system, similar to those found by \citet{2019MNRAS.490..243C} with the uGMRT 300-500 MHz feeds.

\begin{figure}
	\centering
	\includegraphics[width=\columnwidth, trim = {40mm 40mm 40mm 40mm}, clip]{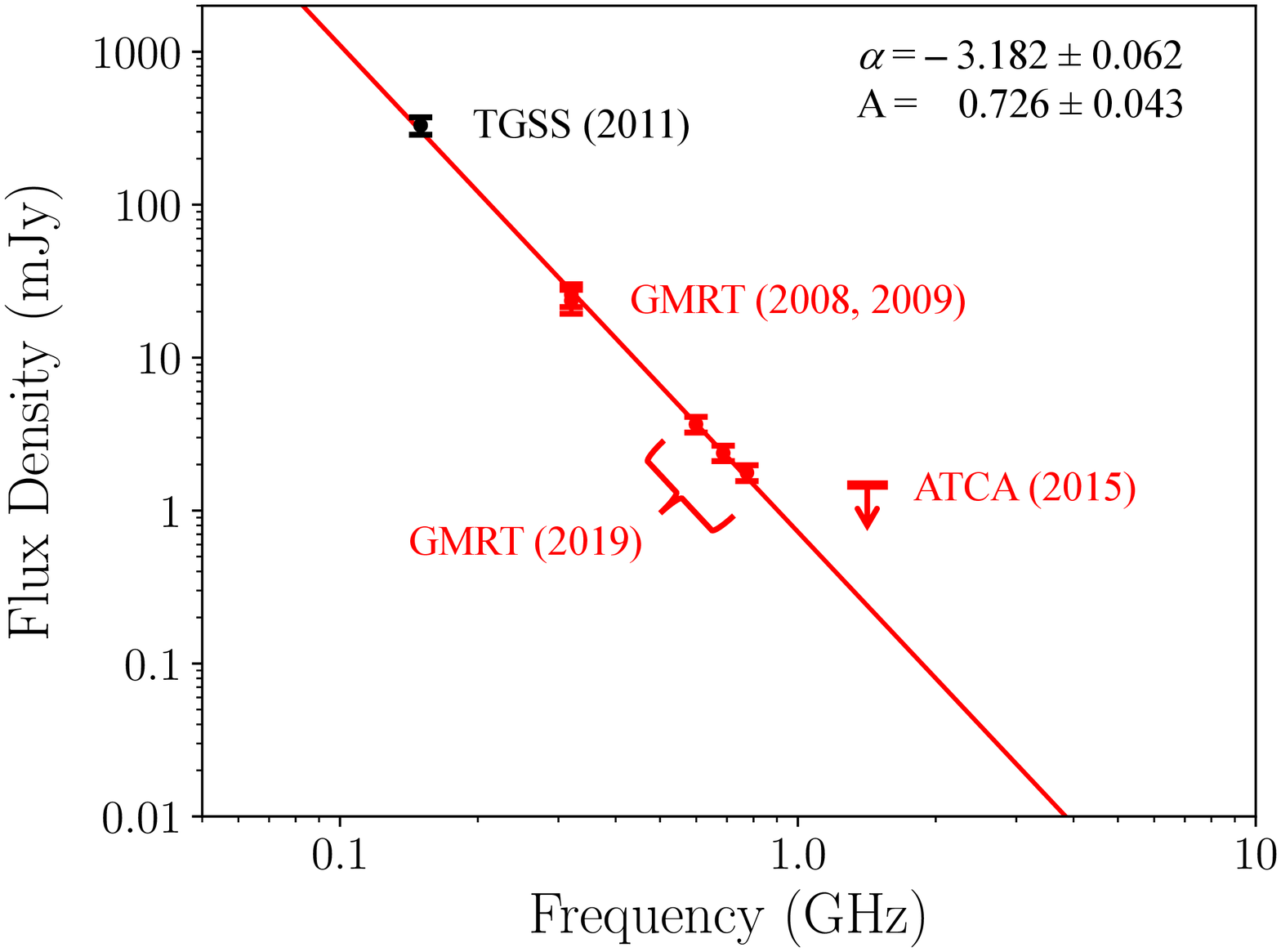}
	\caption{\small Radio spectrum of C1709$-$3918. Results from this work [GMRT (2008, 2009, and 2019) and ATCA (2015)] are shown with red symbols, and results from previously reported data are shown with black symbols.  The red line is a power-law fit using all data presented here, with amplitude, A (in mJy GHz$^{-\alpha}$), and spectral index, $\alpha$, given in the upper right corner. The TGSS (2011) data come from \citet{ijmf16}.}
	 \label{fig:isabel_spectra}
\end{figure}

To check the accuracy of the flux scale, spectra were derived for two bright field sources based on their 150 MHz (TGSS) and 843 MHz \citep{mauch03} flux densities, and those at 1.4 GHz (GPSR \citep{2005AJ....130..586W}, NVSS, and/or our new observations mentioned in \S\ref{Isab Pband}). Comparing the predicted flux densities to the measured ones resulted in an average flux scale correction of 0.9 for each of the images. 
After applying the corrections, the flux densities are $3.67 \pm 0.43$, $2.38 \pm 0.28$, and $1.77 \pm 0.21$ $\mJybm$ at 599, 687, and 774 MHz, respectively, where their uncertainties are dominated by the flux scale correction uncertainty.  
The three flux densities are plotted in Figure~\ref{fig:isabel_spectra} along with the archival 330 MHz detections described in \S\ref{Isab Pband} and the 150 MHz (TGSS) detection. A 1.5 mJy, 5$\sigma$ upper limit at 1.4 GHz is also shown, which we derived using a 0.5 GHz-wide portion of an archival 1--3 GHz observation (29 and 30 January 2015; Program C2872) pointed 7$^{\prime}$ from C1709$-$3918 by the Australia Telescope Compact Array (ATCA). A weighted fit was applied yielding a spectral index of $\alpha$ = $-3.18 \pm 0.06$ for C1709$-$3918.

\subsubsection{C1709$-$3918: Polarization Reduction and Results}\label{isabel1_GMRT_pol}

The polarization calibration and analysis was carried out following the steps described in the VLA 3C\,391 CASA guide\footnote{\url{https://casaguides.nrao.edu/index.php/VLA\_Continuum\_Tutorial\_3C391-CASA5.4.0}}. After initial data flagging, delay, bandpass and gain calibration (see \S\ref{isabel1_GMRT_obs}), we first used {\em setjy} to set the polarized model for the 
position-angle calibrator 3C\,286, for which a 610 MHz fractional polarization of 2.7\% has been reported with the GMRT
\citep{2012PhDT.........4F}.
We then used 3C\,286 to solve for the cross-hand delays and R--L angle. To solve for the instrumental polarization, we first attempted using the 
leakage-term calibrator 1047+384. However, due to significant RFI during the 1047+384 scans and lack of solutions for several antennas,
we used instead the phase calibrator 1714--252 that was observed over a wide enough range of parallactic angles. We found that the average leakage amplitude is within $8\%$, except for 7 antennas that show leakages $>15\%$. These high leakage amplitude antennas were flagged for the polarization analysis \citep[e.g.][]{2014MNRAS.437.3236F}. After further data editing, the final polarization calibrated dataset has a bandwidth 
of 120 MHz and a total of 20 antennas (including 11 out of the 14 antennas in the central compact array). 
We inspected stokes V images of 3C\,286 and 3C\,48 and found no significant detection (circular polarization fraction $<0.9\%$).

Imaging the 120 MHz-wide, polarization calibrated data followed the same procedure as that described in \S\ref{Isab Pband} for the full band. Based on the same two field sources, flux scale correction were applied to three 40 MHz-wide subband images. The Stokes I flux densities of C1709-3918 on each are 4.32 $\pm$ 0.29 $\mJybm$, 3.36 $\pm$ 0.25 $\mJybm$, and 2.94 $\pm$ 0.18 $\mJybm$ for central frequencies of 575 MHz, 614 MHz, and 653 MHz, respectively, consistent with those obtained over the full 270 MHz-wide band.

\begin{figure*}
    \begin{minipage}{\textwidth}
	\centering
	\includegraphics[width=\textwidth, trim = {0mm 35mm 10mm 35mm}, clip]{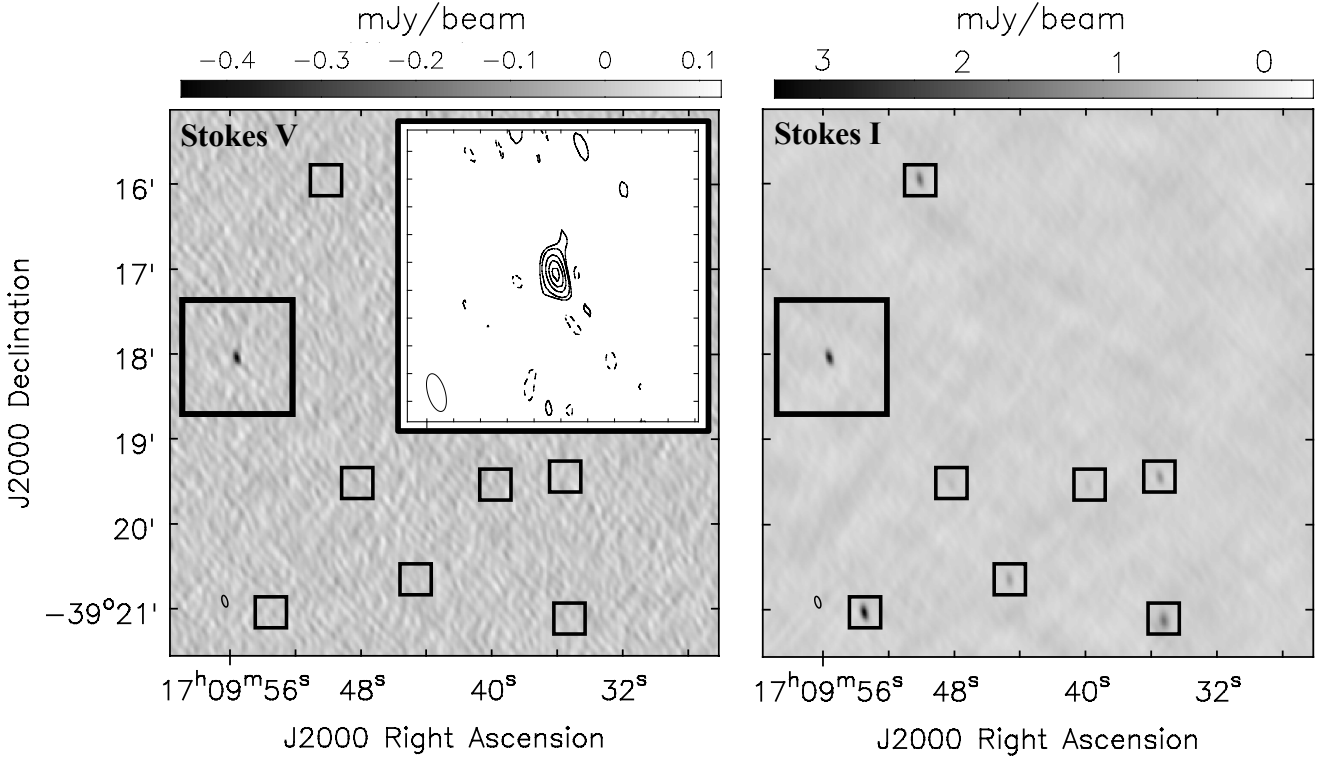}
	\caption{\small Clean map of the circular polarization ({\it left panel}) and total intensity structure ({\it right panel}) of C1709$-$3918 and nearby sources on 2019 June 27. These images are generated using a 120 MHz-wide portion of the bandpass centered at a frequency of 618 MHz. The large black box gives the location for C1709$-$3918. The small black boxes give the locations of nearby sources detecteable in Stokes I, but not Stokes V. The inset image shows the circular polarization intensity (Stokes V) of C1709$-$3918 plotted at $+3$, $-3$, $-5$, $-10$, $-15$, and $-20$ times the rms error, $20\,\mu$Jy beam$^{-1}$ (positive and negative contours are plotted with black dashed and solid lines, respectively).  The resolution of the image is $8.1^{\prime\prime} \times 3.7^{\prime\prime}$ (plotted in the lower-left corner of each image). The fractional circular polarization is $13.8 \pm 0.9\%$.}
	 \label{fig:radio_map_isabel}
    \end{minipage}
\end{figure*}

We then made Stokes I and Stokes V images using the 120 MHz-wide portion of the band. A clear Stokes V detection of -0.44 $\pm$ 0.02 $\mJybm$ is shown in Figure~\ref{fig:radio_map_isabel}, and the corresponding Stokes I flux density is 3.20 $\pm$ 0.09 $\mJybm$, both also listed in Table~\ref{tab:IQUVflux}. Flux scale corrections are not necessary to compute the fraction of circular polarization. We obtain a value of $13.8 \pm 0.9\%$. Dividing the data into 20 MHz-wide subbands yields flux densities of:  $10.5 \pm 2.3\%$ (566 MHz), $11.7 \pm 1.9\%$ (584 MHz), $14.7\pm 2.3\%$ (604 MHz), $16.2 \pm 2.7\%$ (623 MHz), $15.4 \pm 2.4\%$ (643), and $16.0 \pm 1.9\%$ (662 MHz). 

To analyze the linear polarization Stokes parameters Q and U corresponding to C1709$-$3918, we divided the data into two 40 MHz subbands with center frequencies of 596 MHz and 633.5 MHz, respectively. In both subbands, Stokes U was not detectable; however, Stokes Q was detectable: 0.207 $\mJybm$ ($3.6\sigma_Q$) at 596 MHz and 0.237 $\mJybm$ ($5.0\sigma_Q$) at 633.5 MHz. The fractional polarization for the stronger detection in the upper band is Q/I = $7.4\pm1.5\%$.

The error reported here is the rms error determined from the radio maps, primarily due to radiometer noise. A second source of error in measuring the linear polarization may be residual antenna leakage. When we imaged the back-up calibrator 3C\,48, we measured a very low linear polarization of $\approx2\%$. Very little linear polarization information is available for this source, but 3C\,48 is believed to be unpolarized at these frequencies. Assuming that 3C\,48 is unpolarized, this suggests a residual linear polarization leakage of $\approx2\%$ after the manual polarization calibration process. If we add this error in quadrature to the rms error due to radiometer noise, then the fractional polarization for the stronger detection in the upper band is Q/I = $7.4\pm2.5\%$.

\subsection{Pulsation Searches}\label{single}

We observed C1748$-$2827 and C1709$-$3918 with the  Robert C. Byrd Green Bank Telescope (GBT) under a joint Fermi-GBT program (project code AGLST121047). The observations used the VEGAS backend and the 2~GHz (S-band) and 6~GHz (C-band) receivers. VEGAS consists of eight CASPER ROACH2 FPGA boards and eight high performance computers (HPCs). Each FPGA-HPC pair comprises a spectrometer bank.  Our 2~GHz observations used one bank with a bandwidth of 800~MHz, 2048 channels, and a sampling time of 81.92~$\mu$s. Part of the 2~GHz receiver bandwidth is permanently filtered such that the effective bandwidth of our observations is 610~MHz. Our 6~GHz observations used four VEGAS banks, each with a bandwidth of 1500~MHz, 4096 channels, and a sampling time of 87.38~$\mu$s; the bandwidths covered by the four banks are partially overlapped in frequency. Before any further data processing, the data from the four banks were merged into one data set with an effective bandwidth of 4500~MHz and 12288 channels. In addition, Breakthrough Listen is conducting a comprehensive search for evidence of intelligent life near the Galactic Center \citep{BLteam2021} and C1748$-$2827 was observed as a part of that campaign. We used the Breakthrough Listen digital back end \citep{2017PASP..129e4501I,2018PASP..130d4502M} to conduct 60-minutes of deep observations at 6 GHz. This observation had 13312 channels and a sampling time of 44~$\mu$s. Table~\ref{tab:gbt} summarizes all observations. 

We used PRESTO\footnote{\url{http://www.cv.nrao.edu/\~sransom/presto}} \citep{2011ascl.soft07017R} to process the GBT data, excising RFI with {\tt rfifind}, and searching for pulsed signals with {\tt accelsearch}. Most known millisecond pulsars reside in binary systems, where their observed pulse periods are subject to Doppler shift due to orbital motion. The {\tt accelsearch} executable accounts for this by adding an extra dimension to the search space, corresponding to the line-of-sight acceleration of the pulsar relative to the observer. This acceleration parameterizes the Doppler shift. 

Another dimension of the search space is the dispersion measure (DM), which is the integrated column density of ionized gas along the line of sight and is correlated with distance. A distance of 8.5~kpc towards SgrA* corresponds to a DM of $\sim 2500$~pc~cm$^{-3}$ according to the YMW16 model of ionized gas in the Galaxy \citep{ymw17}. We searched the 2~GHz data at 7950 trial DMs in the range $0 - 2687$~pc~cm$^{-3}$, and the 6~GHz data at 2862 trial DMs in the range $0 - 2862$~pc~cm$^{-3}$. The excess above 2500~pc~cm$^{-3}$ in these search parameters is due to technical requirements relating the number of trial DMs to the number of computing nodes and threads used during multi-threaded data processing. (For a more detailed discussion of how trial DMs lists are constructed, see \S{2.4} in \citet{2019ApJ...876...20H}.

No pulsations were detected for either source. The observations and upper limits on the pulsed emission are listed in Table~\ref{tab:gbt}. For calculating the upper limits we used $T_{sky}$ and temperature spectral index values from \citet{Reich90} and \citet{Reich88} to arrive at the $T_{sky}$ values in the table which are scaled to our observing frequencies. 

For C1748$-$2827, a single pulse search yielded no evidence of isolated dispersed pulses above 6$\sigma$ in our earlier 1.5 GHz GBT observations reported in \citet{2019ApJ...876...20H}.
We also searched current GBT observations at 6 GHz for dispersed single pulses from C1748$-$2827 using the SPANDAK pipeline described in \citet{2018ApJ...863....2G}. We searched for single pulses up to the DM of 3000 pc-cm$^{-3}$ with pulse widths ranging from 0.3 to 77 ms. We were able to constrain the presence of isolated 1 ms wide single pulses with a 6$\sigma$ sensitivity of 25 mJy. Additionally, a 72-min integration at 1.5 GHz was re-processed from the High Time Resolution Universe South Low Latitude (HTRU-S LowLat) pulsar survey \citep{2020MNRAS.493.1063C}
in a line of sight near C1709$-$3918, but no pulsations were found (Vishnu Balakrishnan and Rahul Sengar, {\it private communication}).

\begin{table*}
    \begin{minipage}{126mm}
	\centering
	\caption{Pulsation search observations with the GBT. The column at the far right shows upper limits on the pulsed flux density assuming a 10\% pulse duty cycle and a 8$\sigma$ detection threshold.}
	\smallskip
	\begin{threeparttable}
		\begin{tabular}{cccccccccc}
		\hline\hline \noalign{\smallskip} 
        Source &  Date &  $f_c$ &  $\Delta\nu$ &  G &  $T_{sys}$ &  $T_{sky}$ &  $dt$ &  $T_{obs}$ & $S_{UL}$ \\
        & &  (GHz) &  (MHz) &  (K/Jy) &  (K) &  (K) &  ($\mu$s) &  (min) &  ($\mu$Jy)\\
		\hline\noalign{\smallskip}
        2019 Aug 15 & C1709$-$3918 & 2 & 610 & 1.9 & 20 & 10 & 82 & 180 & 12 \\
        2019 Aug 23 & C1709$-$3918 & 2 & 610 & 1.9 & 20 & 10 & 82 & 180 & 12 \\
    \hline
        2019 Nov 27 & C1748$-$2827 & 2 & 610 & 1.9 & 20 & 10 & 82 & 377 & 8 \\
        2019 Jan 11 & C1748$-$2827 & 6 & 4875 & 1.9 & 18 & 1 & 44 & 60 & 5 \\
        2019 Dec 1 & C1748$-$2827 & 6 & 4500 & 1.9 & 18 & 1 & 87 & 400 & 2 \\
		\noalign{\smallskip} 
		\hline\noalign{\smallskip} 		
		\end{tabular}
	\medskip
	\end{threeparttable}
    \label{tab:gbt}
    \end{minipage}
\end{table*}

%

\subsection{X-ray and Optical/NIR search}\label{sec:counterparts}


We searched for X-ray emission within a 10-arcsec radius of the position for each of the two sources using data from the XMM-Newton Science Archive and the Chandra X-ray Center \citep{2017ASPC..512..309E, 2020A&A...641A.137T}. There are no cataloged sources within this regions, and for C1748$-$2827 the estimated Chandra upper limit on the energy flux required for a point source detection is $>5.3\times{10}^{-15}$ erg s$^{-1}$ cm$^{-2}$ (0.5-7.0 keV). Using the HEASARC web-based PIMMS, and assuming a power-law photon index $\Gamma$=$2.0$, the lower limit to the X-ray luminosity for a source in this direction within the bulge is 
4.6$\times{10}^{31}$ erg s$^{-1}$ (2-10 keV).  Likewise, we searched without success for optical/NIR point sources within 0.5$^{\prime\prime}$ of several catalogs including NEOWISE and the VVV survey \citep[][]{mle10,cwc13}.

\subsection{Constraints on Variability}\label{sec:variability}

Each set of observations of C1748$-$2827 and C1709$-$3918 include multiple epochs with similar resolutions and central frequencies. Comparing their flux densities at different epochs yields no evidence of significant variable or transient behavior on any timescale. 

At 330 MHz, the flux density of C1748$-$2827 is consistent to within $\sim20\%$ on year-to-decade timescales from 1998 to 2021, and those for C1709$-$3918 in 2008 and 2009 agree to within their $20\%$ uncertainties. At 1.5 GHz, the two C1748$-$2827 flux densities on 15 and 25 Jan 2021 agree to within $3\%$ (see Table~\ref{tab:IQUVflux}), and with the Jan 2017 observation to within $10\%$. 
For the nine 30 min scans of C1709$-$3918 at 618 MHz, the ratio of the standard deviation to average fluctuation is 14\%, with the largest deviation being 26\%. For seventeen 15 min segments of the two C1748$-$2827 observations at 1.5 GHz, the ratio is 0.09, with the largest deviation being 21\%.



\section{Discussion and Conclusions}\label{sec:conclusions}

 We have refined the astrophysical parameters of two steep-spectrum radio sources, from a larger sample of nearly two dozen discovered toward the bulge of our Galaxy. Their full properties are listed in Table \ref{tab:psr}, but summarizing, C1748$-$2827 and C1709$-$3918 are both compact, and steep spectrum. Their flux densities are relatively steady on timescales of years, months and hours, with no indications of rapid variability or transient behaviors. They are {\it circularly} polarized, but their fractional {\it linear} polarizations are either low or undetected. They lack counterparts, at both X-ray and optical/NIR wavelengths, and, despite deep searches up to 6~GHz, we find no evidence for pulsations. We next compare the properties of C1748$-$2827 and C1709$-$3918 to those of known compact non-thermal radio source populations, with the goal of better understanding their astrophysical nature. We show that the most likely hypothesis, consistent with the existing data, is that these are likely scatter-broadened pulsars.
 
X-ray binaries (XRBs; consisting of BH and NS systems) are a common population toward the bulge \citep{2006ApJS..165..173M,2018Natur.556...70H}. The radio luminosities of both sources (Table \ref{tab:psr}) are consistent with the range observed for BH systems, but at least an order of magnitude larger than that observed for NS binaries \citep[see][]{2020ApJ...905..173Z}.  However, the radio counterparts of XRBs tend to have flat spectra and have considerable variability, approaching 100\% with outbursts then fading into quiescence.  These latter two properties are inconsistent with both sources. 

CVs are a common population toward the inner Galaxy.  Radio emission has been detected from CVs, including circularly polarized emission, but CVs are both variable and have luminosities at least 3 orders of magnitude fainter than these sources \citep{2020AdSpR..66.1226B,2020NewAR..8901540C}. 


\begin{table*}
    \begin{minipage}{126mm}
	\centering
	\caption{Derived Source Parameters.}
	\smallskip
	\begin{threeparttable}
		\begin{tabular}{lll}
		\hline\hline \noalign{\smallskip} 
		Parameter & C1748$-$2827 & C1709$-$3918\\
		\hline\noalign{\smallskip}
        Right Ascension (J2000) & 17${\rm ^h}$ 48${\rm ^m}$ 07${\rm ^s}$.070(15) & 17${\rm ^h}$ 09${\rm ^m}$ 55${\rm ^s}$.570(35)\\
        Declination (J2000) & $-$28$^\circ27^\prime41.68\left(10\right)^{\prime\prime}$ & $-$39$^\circ18^\prime02.05\left(61\right)^{\prime\prime}$\\
        Galactic Longitude & 0.6899$^\circ$   &  347.2926$^\circ$\\
        Galactic Latitude  & $-$0.2220$^\circ$  & 0.3641$^\circ$\\
        Angular Diameter (arcsec) & $<$0.9 & $<$2.0 \\
        Spectral index $\alpha$ ($S \propto \nu^\alpha$) & $-$2.85$\pm$0.03& $-$3.18$\pm$0.06 \\
        Linear Polarization (\%) & $<$7.6 & 7.4$\pm$2.5 \\
        Circular Polarization (\%) & 14.4$\pm$1.5 & 13.8$\pm$0.9 \\
        Variability (\%) & 3-20 & 9-20 \\
        Radio Luminosity $L_{\rm R}$\tnote{a,b}\hspace{3mm} (erg/s) & 3.5$\times{10}^{30}$ & 3.9$\times{10}^{30}$ \\
        Pseudo-luminosity $L_{\rm 1.4GHz}$\tnote{b}\hspace{2mm} (mJy~kpc$^{2}$) & 29 & 16 \\
		\noalign{\smallskip} 
		\hline\noalign{\smallskip} 		
		\end{tabular}
	\medskip
		\begin{tablenotes}
		\footnotesize
		\item[a] Integrated power-law spectra from 100 MHz to 10 GHz.
        \item[b] Adopted a typical bulge distance d 8~kpc.
		\end{tablenotes}	
	\end{threeparttable}
    \label{tab:psr}
    \end{minipage}
\end{table*}

Both sources share some similarities with GCRT\,J1745$-$3009 (a.{}k.{}a.\ ``The Burper''), i.e., high circular polarization, a steep spectrum, and the absence of quiescent counterpart at other wavelengths \citep{2010ApJ...712L...5R, 2009A&A...502..549S,Hyman05}. The similarities diverge from there, however, as GCRT\,J1745$-$3009 appears to be a unique transient \citep{2016ApJ...832...60P,2019ApJ...886..123A}, discovered initially as five Jansky-like flares of 10-min duration with a 77-min recurrence timescale \citep{Hyman05}, followed by some additional bursts before fading away altogether \citep{2006ApJ...639..348H, 2007ApJ...660L.121H}. Another unique GC source is magnetar PSR\,J1745$-$2900, and although it is polarized like the sources here, it pulsates at a known period of 3.76 s, and is time-variable in both X-rays, and radio and has a flat radio spectrum \citep{2013MNRAS.435L..29S,2015ApJ...806..266L,2018ApJ...866..160P}.

The source populations that most closely match the properties of C1748$-$2827 and C1709$-$3918 are normal pulsars and the recycled or millisecond pulsars (MSPs).  
In a comparison of 31 MSPs and 369 normal pulsars, \citet{1998ApJ...501..270K} found that MSPs are approximately ten times fainter than normal pulsars with $\mathrm{L}_{1.4}$ = $3 \pm 2$~mJy~kpc${}^2$ versus $\mathrm{L}_{1.4}$ = $32 \pm 3$~mJy~kpc${}^2$, respectively.
While the 1.4~GHz luminosities of C1748$-$2827 and C1709$-$3918 are typical of normal pulsars (Table~\ref{tab:psr}), they would be considered luminous outliers if they are MSPs. Likewise, their spectral indices are in the (extremely) steep tail for most known pulsars. In the most recent ATNF Pulsar Catalog \citep{2005AJ....129.1993M} less than 2\% of all pulsars with well-determined spectral indices have $\alpha<-1.8$. The mean spectral index of MSPs is $\overline{\alpha} $=$ -1.76\pm{0.01}$ \citep{2015MNRAS.449.3223D}, similar to normal pulsars \citep{blv13,mkk00}. Neither of these discrepant properties by themselves should be sufficient to rule out a pulsar origin, as one would expect that a flux-limited, image-based search such as this one to be biased toward luminous pulsars \citep[e.g.][]{1995ApJ...455L..55N}.

Pulsars are also polarized. The {\it phase-resolved} polarized emission from normal and MSPs can be very different from each other, with MSPs typically having more complex profiles \citep{1998ApJ...501..286X}. However, their {\it time-averaged} polarization properties are similar \citep{2015MNRAS.449.3223D,2021arXiv210402294B} and the approximately 15\% fractional circular polarizations seen for C1748$-$2827 and C1709$-$3918 is not unusual. The weak or absent {\it linear} polarization signal is unusual and could suggest different emission mechanisms such as gyro-synchrotron, where circular polarization dominates. Alternatively, the fractional linear polarization could be reduced by bandwidth depolarization if there is a strong magneto-ionized medium along the lines of sight, without affecting the circular polarization. This will reduce the polarization by a factor of $\sin{(\psi_{\mathrm{RM}})}$/$\psi_{\mathrm{RM}}$, where
\begin{equation}
\psi_{\mathrm{RM}} \equiv  -\left(\frac{2\Delta\nu}{\nu}\right)\mathrm{RM}\lambda^2
\end{equation}
and $\nu$, $\lambda$, $\Delta\nu$ are the center frequency, corresponding wavelength, and bandwidth of the observation, respectively; RM is the Faraday rotation measure of the magneto-ionized medium \citep[equation~24]{Gardner&Whiteoak:1966}. For $\psi_{\mathrm{RM}} \approx 1$ rad, the polarization decreases to $\approx84\%$ of the intrinsic value; similarly, for $\psi_{\mathrm{RM}}$ = $\pi$ rad, the source becomes completely depolarized. Table~\ref{tab:BW_depol} shows the maximum bandwidth allowed for a range of rotation measures corresponding to typical values of RM within $1\,\mathrm{kpc}$ of the Galactic center \citep[e.g. see][]{2021MNRAS.502.3814L}, assuming $\psi_{\mathrm{RM}} \approx 1$ rad and $\approx \pi$ rad in columns four and five, respectively.



\begin{table}
	\centering
	\caption{The maximum bandwidth ($\Delta\nu$) allowed for a given observation frequency ($\nu$) and Faraday rotation measure (RM). $\Delta\nu$(1) and $\Delta\nu$($\pi$) are the bandwidths at which linear polarization decreases to $84\%$ and $0\%$, respectively, of the intrinsic value due to bandwidth depolarization.}
	\smallskip
	\begin{threeparttable}
		\begin{tabular}{lcccc}
		\hline\hline \noalign{\smallskip} 
        \multirow{2}{*}{Source} & $\nu$ & RM  & $\Delta\nu(1)$ & $\Delta\nu(\pi)$\\
         & (MHz) & (rad m$^{-2}$) &(MHz) &  (MHz)\\
		\hline\noalign{\smallskip}
        \multirow{4}{*}{C1748$-$2827} & \multirow{4}{*}{1267}  & 50 &226&711\\
         & & 100 &113&355\\
         & & 1000 &11.3&35.5\\
         & & 2000 &5.7&17.8\\
     \hline
        \multirow{4}{*}{C1709$-$3918} & \multirow{4}{*}{633.5} & 50 & 28 &89\\
         & & 100 & 14&44\\
         & & 1000 & 1.4&4.4\\
         & & 2000 &0.7&2.2\\
		\noalign{\smallskip} 
		\hline\noalign{\smallskip} 		
		\end{tabular}
	\end{threeparttable}
    \label{tab:BW_depol}
\end{table}

The smallest bandwidth analyzed for C1748$-$2827 was 56 MHz. In all frequency bands analyzed (spw $0-7$), there was no clear detection of Q or U. If we attribute this to bandwidth depolarization, then the data for C1748$-$2827 place a lower limit on the Faraday rotation along the line of sight: $|\mathrm{RM}|\ge634$ rad m$^{-2}$. By comparison, \cite{2021MNRAS.502.3814L} determined the RM of 62 sources within $1\,\mathrm{kpc}$ of the Galactic center. For the lines of sight at the same low Galactic latitude of our two sources ($\vert{b}\vert\leq1^\circ$), the RM reported in Table 1 of \cite{2021MNRAS.502.3814L} include $593.4\leq\mathrm{RM}\leq1691.2$, with an average value of
$+1040.0$ rad m$^{-2}$ and standard deviation of $348.1$ rad m$^{-2}$. The RM limit for C1748$-$2827 is consistent with this range. It is larger than the smallest RM, but still $\approx1000$ rad m$^{-2}$ less than the maximum value. If the Faraday rotation along the lines of sight to each source is between $1000-2000$ rad m$^{-2}$, then Table~\ref{tab:BW_depol} demonstrates that to clearly detect linear polarization, bandwidths of $\le5$ MHz are required.

In the case of C1709$-$3918, a weak detection of Stokes Q was made at the $5.0\sigma_Q$ level at 633.5 MHz, using a bandwidth of 40 MHz. If we assume this detection is real, then if the true fractional Q/I is $10\%-25\%$, bandwidth depolarization reduction to $7.4\%$ corresponds to an RM in the range of $46-84$ rad m$^{-2}$. If we, instead, assume that this was a spurious detection, then we can use these data to derive a lower limit of $|\mathrm{RM}|\ge111$ rad m$^{-2}$. Because the GMRT observations of C1709$-$3918 are at lower frequencies, the necessary bandwidth required to detect an RM of $1000-2000$ rad m$^{-2}$ is even more restrictive: $\approx1$ MHz, which is the native frequency resolution of the GMRT. Thus we can understand the absence of significant linear polarization from 
C1748$-$2827 and C1709$-$3918 as the result of their lines of sight having to propagate through a large Faraday depth, consistent with a bulge distance. Another weak distance constraint is the absence of significant variability (\S\ref{sec:variability}), as it is well-known that pulsars with higher DMs (i.e., larger distances) have less variable flux densities \citep[see Fig. 5 of][]{2000ApJ...539..300S}.

Pulsations are a defining characteristic of pulsars, and their absence needs to be explained. For C1709$-$3918 we have searched for pulsations without success up to 2 GHz, while for C1748$-$2827 we have verified previous null results up to 2 GHz \citep{2019ApJ...876...20H} and further extended the deep searches to 6 GHz. There was sufficient sensitivity to detect pulsations as can be seen comparing the limits in Table \ref{tab:gbt} to the spectra in Figs.\,\ref{fig:platinum_spectra} and \ref{fig:isabel_spectra}. The simplest hypothesis is that C1748$-$2827 and C1709$-$3918 are indeed pulsars but their pulses are scattered broadened by the excess turbulent ionized gas along the line of sight toward the Galactic center. The main uncertainty in testing this hypothesis lies with calculating the magnitude of pulse smearing. The temporal smearing estimates $\tau_{scat}$ near the GC from the two most common models \citep{cl02,ymw17} differ by nearly two orders of magnitude, owing to the different methodologies used. The lower scattering estimates from \citet{ymw17} are supported by the observed angular and temporal broadening toward the magnetar PSR\,J1745$-$2900 \citep{2014ApJ...780L...2B,2014ApJ...780L...3S}. In \citet{2019ApJ...876...20H} we showed for either model, estimates of $\tau_{scat}$ at 2 GHz made it difficult to ``hide'' a normal pulsar along the line of sight to the bulge (d $<$ 8 kpc), based on interstellar scattering alone. This conclusion is further bolstered by the new 6 GHz limits toward C1748$-$2827, since $\tau_{scat}$ scales strongly with inverse frequency (i.e. $\nu^{-4}$). In the weak scattering models of \citet{ymw17} and \citet{2014ApJ...780L...3S} $\tau_{scat}$ at 6 GHz at GC distances are reduced to 7 ms and 1 ms, respectively, sufficiently small to detect normal pulsars and most known MSPs. In the strong scattering model of \citet{cl02} the temporal broadening $\tau_{scat}$ = 0.7 s, making our search sensitive only to magnetar-like periods. For C1709$-$3918, which is 12.7$^\circ$ from the enhanced scattering around SgrA*, the 2 GHz model-estimated values of $\tau_{scat}$ at the distance of the bulge are approximately 3.7 ms \citep{cl02} and 19 ms \citep{ymw17}, respectively. If C1748$-$2827 and C1709$-$3918 are pulsars, the phase space available for them to remain undetected is shrinking.

If history is a guide, pulsations will eventually be detected from C1748$-$2827, C1709$-$3918, and some of the other bulge candidates. The compact, steep spectrum, polarized radio sources, identified toward 4C\,21.53 and M28 \citep{1979ApJ...228..755R,1980BAAS...12..799E,1985AJ.....90..606H,1987ApJ...314L..45E} led to the discovery of the first isolated millisecond pulsar PSR\,B1937+21 \citep{1982Natur.300..615B} and the first globular cluster millisecond pulsar PSR\,B1821$-$24 \citep{1987Natur.328..399L}, respectively. Small numbers of radio pulsars continue to be discovered by targeting pulsations searches toward compact, steep spectrum (polarized) radio sources identified through interferometric imaging \citep{1995ApJ...455L..55N,1987ApJ...319L.103S,2017MNRAS.468.2526B,2018MNRAS.475..942F,2019ApJ...884...96K}. Increasingly however, these image-based surveys are finding large numbers of compelling candidates \citep{2000ApJ...529..859K,2000AJ....119.2376C,2000A&AS..143..303D,2018MNRAS.474.5008D} in which pulsation searches have been unsuccessful \citep{2012A&A...546A..25R,2013AJ....145..116S,2018ApJ...864...16M,2019ApJ...876...20H,2019A&A...623A..90S,2021RNAAS...5...21C}. While there are prosaic explanations for why pulsations might not have be seen
\citep[see][]{2018MNRAS.474.5008D,2018ApJ...864...16M}, the sample size of non-detections is growing large. C1748$-$2827 and C1709$-$3918 are just the latest and best-studied examples and they raise the possibility, first suggested by \citet{2018ApJ...864...16M}, that they represent a new population of Galactic radio sources which share similar properties as pulsars but lack pulsations. Future large-scale synoptic surveys which plan to use compactness, spectra and/or polarization to identify potential pulsar candidates \citep[e.g.]{2017MNRAS.472.1458D,2018MNRAS.478.2835L} are well-suited to test the likelihood of this hypothesis.

The present study, while an improvement over early work, still has a number of shortcomings. While we have identified nearly two dozen steep spectrum candidates in the bulge, most have not received the same level of attention as C1748$-$2827 and C1709$-$3918. Thus broad statements about putative source properties of a bulge population should be taken with caution. More importantly, we lack any significant constraints on the distances to these sources. In population synthesis modeling of pulsars in the bulge, \citet{2016ApJ...827..143C} showed that disk MSPs were a significant foreground contaminant to bulge searches. Likewise, some of these compact radio sources could be background AGN, a rare type of steep-spectrum, luminous high redshift galaxy \citep{2008A&ARv..15...67M}. Estimates of the amount of ionized gas along the line of sight could serve as a distance indicator. We have argued above that the weak or undetected fractional {\it linear} polarization is suggestive of a far distance. Further constraints at MHz frequencies on the low frequency end of the spectra of these sources could be used to estimate the amount of free-free absorption. These observations assume that the spectral turnover and low polarization have an extrinsic origin, and are not intrinsic properties; an assumption that may not be well-justified in practice.
The most promising future option appears to be milliarcsecond imaging of the steep spectrum candidates. Sources within the bulge are expected to show significant angular broadening $\theta_{\mathrm{scat}} \propto \nu^{-2}$.
Indeed, the large angular diameters measured for the magnetar PSR\,J1745$-$2900 \citep{2014ApJ...780L...2B} and the X-ray binary  GCT \citep{1992Sci...255.1538Z} were used to argue that they were at the same distance as Sgr A*. With such imaging it may be possible to resolve background AGN and identify unresolved foreground pulsars, leaving a sample of bulge candidates for further detailed searches.

\section*{Acknowledgements}

We thank the Director of the VLA at the National Radio Astronomy Observatory, Dr. Mark McKinnon, for granting Director's Discretionary Time to search for polarized emission from C1748$-$2827.
We thank Scott Ransom for revisiting one of our GBT observations for an independent pulsation search, and the HTRU collaboration for reprocessing the HTRU-S LowLat data.
We thank the staff of the GMRT that made observations possible. GMRT is run by the National Centre for
Astrophysics of the Tata Institute of Fundamental Research. The National Radio Astronomy Observatory is a facility of the National Science Foundation operated under cooperative agreement by Associated Universities, Inc. The Green Bank Observatory is a facility of the National Science Foundation operated under cooperative agreement by Associated Universities, Inc. Portions of this work at NRL and SBC were funded by NASA. Basic research in radio astronomy at NRL is supported by the Chief of Naval Research (CNR). Part of this research was carried out at the Jet Propulsion Laboratory, California Institute of Technology, under a contract with the National Aeronautics and Space Administration. The NANOGrav project receives support from National Science Foundation (NSF) Physics Frontiers Center award number 1430284. This research has made use of data and/or software provided by the High Energy Astrophysics Science Archive Research Center (HEASARC), which is a service of the Astrophysics Science Division at NASA/GSFC. This research has made use of the SIMBAD database, operated at CDS, Strasbourg, France. This research has made use of the NASA/IPAC Extragalactic Database (NED) which is operated by the Jet Propulsion Laboratory, California Institute of Technology, under contract with the National Aeronautics and Space Administration. Breakthrough Listen is managed by the Breakthrough Initiatives, sponsored by the Breakthrough Prize Foundation.

\section*{Facilities}
Karl G.~Jansky Very Large Array (VLA), Green Bank Telescope (GBT), Giant Metrewave Radio Telescope (GMRT)


\section*{Software}
AIPS (31DEC19, 31DEC18), CASA (5.6.1, 5.4.1, 5.1.0), OBIT, PRESTO





\section*{Data Availability}

The data underlying this article were accessed from the GMRT (Project 36044; \url{https://naps.ncra.tifr.res.in/goa/data/search}) and VLA (Project 20B-461; \url{https://science.nrao.edu/facilities/vla/archive/index}) data archives. The derived data generated in this research will be shared on reasonable request to the corresponding author.

{}

\label{lastpage}
\end{document}